\documentclass[twocolumn,showpacs,preprintnumbers,amsmath,amssymb]{revtex4}

\usepackage{graphicx}
\usepackage{dcolumn}
\usepackage{bm}

\begin{document}

\def\piminuspiplus{1.025}
\def\piminuspiplusstaterr{0.006}
\def\piminuspiplussyserr{0.018}
\def\kminuskplus{0.95}
\def\kminuskplusstaterr{0.03}
\def\kminuskplussyserr{0.03}
\def\pbarp{0.73}
\def\pbarpstaterr{0.02}
\def\pbarpsyserr{0.03}

\def\npart{317}
\def\nparterr{10}

\def\abcorrgheisha{5.0}
\def\abcorrfluka{2.1}
\def\abcorr{3.5}
\def\aberr{1.5}


\preprint{APS/123-??}

\title{Centrality dependence of charged antiparticle to particle ratios \\ near mid-rapidity in d+Au collisions at $\sqrt{s_{_{NN}}}=200$ GeV}
\author{	
B.B.Back$^1$,
M.D.Baker$^2$,
M.Ballintijn$^4$,
D.S.Barton$^2$,
B.Becker$^2$,
R.R.Betts$^6$,
A.A.Bickley$^7$,
R.Bindel$^7$,
W.Busza$^4$,
A.Carroll$^2$,
M.P.Decowski$^4$,
E.Garc\'{\i}a$^6$,
T.Gburek$^3$,
N.George$^{1,2}$,
K.Gulbrandsen$^4$,
S.Gushue$^2$,
C.Halliwell$^6$,
J.Hamblen$^8$,
A.S.Harrington$^8$,
C.Henderson$^4$,
D.J.Hofman$^6$,
R.S.Hollis$^6$,
R.Ho\l y\'{n}ski$^3$,
B.Holzman$^2$,
A.Iordanova$^6$,
E.Johnson$^8$,
J.L.Kane$^4$,
N.Khan$^8$,
P.Kulinich$^4$,
C.M.Kuo$^5$,
J.W.Lee$^4$,
W.T.Lin$^5$,
S.Manly$^8$,
A.C.Mignerey$^7$,
R.Nouicer$^{2,6}$,
A.Olszewski$^3$,
R.Pak$^2$,
I.C.Park$^8$,
H.Pernegger$^4$,
C.Reed$^4$,
C.Roland$^4$,
G.Roland$^4$,
J.Sagerer$^6$,
P.Sarin$^4$,
I.Sedykh$^2$,
W.Skulski$^8$,
C.E.Smith$^6$,
P.Steinberg$^2$,
G.S.F.Stephans$^4$,
A.Sukhanov$^2$,
M.B.Tonjes$^7$,
A.Trzupek$^3$,
C.Vale$^4$,
G.J.van~Nieuwenhuizen$^4$,
R.Verdier$^4$,
G.I.Veres$^4$,
F.L.H.Wolfs$^8$,
B.Wosiek$^3$,
K.Wo\'{z}niak$^3$,
B.Wys\l ouch$^4$,
J.Zhang$^4$\\
(PHOBOS Collaboration)\\
\vspace{3mm}
\small
$^1$~Argonne National Laboratory, Argonne, IL 60439-4843, USA\\
$^2$~Brookhaven National Laboratory, Upton, NY 11973-5000, USA\\
$^3$~Institute of Nuclear Physics, Krak\'{o}w, Poland\\
$^4$~Massachusetts Institute of Technology, Cambridge, MA 02139-4307, USA\\
$^5$~National Central University, Chung-Li, Taiwan\\
$^6$~University of Illinois at Chicago, Chicago, IL 60607-7059, USA\\
$^7$~University of Maryland, College Park, MD 20742, USA\\
$^8$~University of Rochester, Rochester, NY 14627, USA\\
}

\date{September 16, 2003}
\begin{abstract}\noindent
  The ratios of the yields of charged antiparticles to particles have been obtained for pions, kaons, and protons near mid-rapidity for d+Au collisions at $\sqrt{s_{_{NN}}} = 200$~GeV as a function of centrality. The reported values represent the ratio of the yields averaged over the rapidity range of $0.1<y_{\pi}<1.3$ and $0<y_{K,p}<0.8$, where positive rapidity is in the deuteron direction, and for transverse momenta $0.1<p_{T}^{\pi,K}<1.0$~GeV/c and $0.3<p_{T}^{p}<1.0$~GeV/c. Within the uncertainties, a lack of centrality dependence is observed in all three ratios. The data are compared to results from other systems and model calculations.
\end{abstract}


\maketitle

Experiments at the Relativistic Heavy Ion Collider (RHIC) at Brookhaven National Laboratory aim to understand the behavior of strongly interacting matter at high temperature and density, testing predictions of quantum chromodynamics (QCD).
As part of this investigation, smaller systems at RHIC energies need to be studied in order to aid in the understanding and interpretation of results from the more complicated heavy ion collisions.
In this paper, the ratios of the yields of antiparticles to particles for primary charged pions, kaons, and protons in d+Au collisions at $\sqrt{s_{_{NN}}} = 200$~GeV as a function of collision centrality were determined using the PHOBOS detector during the 2003 run. 

Anti-proton to proton yield ratios near mid-rapidity depend largely on the dynamics of baryon-antibaryon pair production and baryon number transport in nuclear collisions. The rate of pair production can depend on the state of the matter created, see references in \cite{star130_pbar}. Recent comparisons of d+Au and Au+Au data suggest that the conditions in Au+Au collisions are very different from those observed in d+Au \cite{phobosdAhighpt,phenixdAhighpt,stardAhighpt,brahmsdAhighpt}. Whether these different conditions influence the particle ratios is explored by measuring the ratios in d+Au collisions. 

In Au+Au collisions at RHIC energies the ratio of $\langle \bar{p} \rangle / \langle p \rangle$  increases from $0.6$ to $0.8$ as the collision energy increases from 130~GeV to 200~GeV, and shows a weak dependence on centrality and $p_{T}$ \cite{star130,phobos130,brahms130,phenix130,brahms200,phobos200,phobos200_QM02,star200,phenixspectra}. These results imply that baryon-antibaryon pair production is larger than baryon number transport and yet there is still a finite baryon number transport over 5 units of rapidity~\cite{star130_pbar}.  

Based on lower energy data, the expectation is that the more collisions ($\nu$) a participating nucleon suffers, the greater the baryon number transport to mid-rapidity \cite{wit,wit2}.  This results in the reduction of the $\langle \bar{p} \rangle / \langle p \rangle$ ratio. For central Au+Au collisions (12\% most central events \cite{phobos200}) each participating nucleon suffers on average 5.2 collisions, $\nu \equiv \frac{N_{coll}}{N_{part}/2}$, where $N_{coll}$ and $N_{part}$ are the number of binary collisions and the number of participants, respectively. In d+Au collisions, when looking in the deuteron hemisphere, $\nu$ can be defined using the number of deuteron participating nucleons ($\nu \equiv \frac{N_{coll}}{N_{part}^{d}}$). Over the range of centrality discussed in this paper, $\left<\nu\right>$ varies from 2 to 8. Hence the range of $\nu$ in this d+Au measurement includes that observed in Au+Au collisions.  This allows a comparison of the relative magnitude of the baryon number transport per produced baryon between the two systems. 

Results from Au+Au collisions \cite{brahms200,phobos200} show that the $\langle K^- \rangle / \langle K^+ \rangle$ and $\langle \bar{p} \rangle / \langle p \rangle$ ratios are consistent with thermal models.  This implies that frequent final state interactions occur.  In d+Au collisions little reinteraction is expected and therefore the particle ratios reflect the initially produced yields.  Thus, the influence of baryon number transport, baryon production, and final state interactions can be investigated by comparing results from the two systems.  
 
The results reported in this paper were obtained using the PHOBOS two-arm magnetic spectrometer \cite{phobosNIM}.  
Each arm has a total of 16 layers of silicon sensors, providing charged-particle tracking both outside and inside the 2 T field of the PHOBOS magnet. 
Particles within the geometrical acceptance region used in this analysis traverse at least 12 of the 16 layers.
Three single layer silicon pad detectors (``Ring counters'') located on either side of the interaction point were used to determine the multiplicity in the pseudo-rapidity range covering $3 < \left| \eta \right| < 5.4$. 

Another single layer silicon pad detector (``OCT'') surrounds the interaction region with a cylindrical geometry along the longitudinal (z) direction with $\left| z \right|<50$~cm, corresponding to a pseudo-rapidity coverage of $\left| \eta \right|<3.2$. The z position of the vertex is found by maximizing the number of OCT hits above a variable threshold.  Due to the changing angle of incidence, the energy deposited per track passing through the OCT detector, and hence the low energy cutoff that defines a hit, increases as the distance from the vertex increases.  The resulting vertex position resolution ranges from 0.7~cm to 1.3~cm in central and peripheral collisions, respectively. This method of vertex reconstruction was found to be the most efficient for low-multiplicity events.  

An event-by-event reconstruction of the transverse position of the interaction vertex is not possible due to the low track multiplicity in d+Au collisions.  Instead, the average transverse position of the vertex (beam orbit) for a given data taking period is used. The beam orbit is determined from the intersection points of tracks traversing multiple layers of silicon in the spectrometer planes that lie outside of the magnetic field and in a two layer silicon pad detector covering $|\eta| < 1.5$ and 25\% of the azimuthal angle.

The primary event trigger was provided by two sets of 10 \v Cerenkov detectors (``T0s''), which cover the pseudo-rapidity range $-4.9<\eta<-4.4$~(T0N) and $3.7<\eta<4.2$~(T0P) for the nominal vertex position ($z=0$), where positive $\eta$ is defined as the direction of the deuteron. These asymmetric positions were chosen to optimize the acceptance for primaries, without shadowing the Ring counters. A triggered event required a coincidence between T0N and T0P as well as a time agreement that corresponded to an approximate vertex range of $\left| z \right| < 50$~cm. The data set of 30 million triggered events requiring only this condition is referred to as ``dAVertex''. To enhance the sample of peripheral events, a separate data set with an additional online trigger condition was taken. This trigger required that the occupancy in each of the two sets of 16 scintillator paddle counters, which cover the pseudo-rapidity range $3 < \left| \eta \right| < 4.5$, be less than 50\%. This data set of 20 million triggered events is referred to as ``dAPeriph''.

Offline event selection cuts were applied to both d+Au data sets. To ensure that a non-spurious vertex was reconstructed, a more restrictive cut on the time difference between the T0s was applied. In addition, a cut was applied requiring agreement between the standard and T0 vertices. To achieve uniformity across all PHOBOS trigger configurations, it was required that both sets of paddle counters have at least one hit. To reduce vertex-position-dependent systematic effects, only events with a vertex of $|z|<$~8~cm were used. This range was chosen to ensure that both particles and antiparticles can be tracked and identified in the spectrometer for both polarity settings.

For this analysis, the events were divided into four centrality classes based on the observed total angle-corrected energy deposited ($E_{Ring}$) in the Ring counters, which is proportional to the number of charged particles hitting these detectors. The four centrality classes were determined by cuts in $E_{Ring}$ which correspond to a percentage range of the full $E_{Ring}$ distribution not biased by trigger or vertex inefficiencies. These cuts were determined from a Glauber model calculation using HIJING \cite{hijing} and a GEANT~3.21 simulation of the full detector. Table~\ref{table2} shows the percentages that define these classes of events. The relative multiplicity of each data set to that in the most central 10\% bin ($\left<E_{Ring}^{norm}\right>$) is also shown to provide a model-independent measure of the centrality. Using the HIJING model to relate $E_{Ring}$ and $N_{coll}$, the average $\langle N_{coll} \rangle$, impact parameter $\langle b \rangle$ and number of collisions per deuteron participant $\langle \nu \rangle$ for each trigger condition and centrality bin can be determined. In addition, the average trigger and vertex efficiency $\left<\epsilon\right>$ as determined using HIJING is quoted.  The additional trigger requirements in the dAPeriph data set remove more central events, resulting in a reduced efficiency and also slightly different average values in the 30-60\% centrality bin relative to the dAVertex data set.  

The tracking used in this analysis is similar to that used in previous Au+Au analyses \cite{phobos130,phobos200,phobos200highpt}, and is the same as in \cite{phobosdAhighpt}. For the low-multiplicity environment of d+Au collisions the track seeds can be determined without knowledge of the vertex position. This ``vertexless'' tracking is the major difference between the d+Au and Au+Au tracking algorithms. The tracking was also extended to include additional spectrometer sensors within the field region, as compared to the previous ratio analyses \cite{phobos130,phobos200}. This results in an increased rapidity coverage. A three sigma cut on the track's distance of closest approach to the beam orbit (dca$<0.35$~cm) was applied to reduce the secondary contribution, which is largest at low $p_{T}$.  Additional $p_{T}$ cuts of $0.1$~GeV/c for pions and kaons and $0.3$~GeV/c for protons were applied to keep the secondary contribution below $1\%$ and $5\%$, respectively.

\begin{figure}
\centerline{
\resizebox{0.5\textwidth}{0.25\textwidth}{
\includegraphics{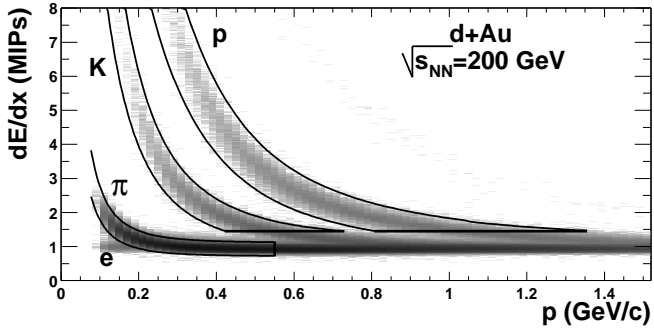}}
}
\caption{Distribution of average truncated energy loss as a function of 
reconstructed particle momentum. Three clear bands can be seen,
corresponding to pions, kaons, and protons. The solid lines indicate
the cuts used for particle identification.}
\label{dedx_p}
\end{figure}
Particle identification (PID) was based on the truncated mean of the specific ionization $dE/dx$ measured in the silicon spectrometer planes \cite{phobos130}. The PID cuts for pions, kaons, and protons are shown in Fig.~\ref{dedx_p}. The curved bands are based on the position where the $\left<dE/dx\right>$ distribution for a given momentum is three RMS deviations away from the mean expected value for each species. The upper momentum cut for the pions and the lower $dE/dx$ cuts for the kaons and protons are determined in order to minimize possible contamination from other particle species.

\begin{figure}
\centerline{
\resizebox{0.5\textwidth}{0.25\textwidth}{
\includegraphics{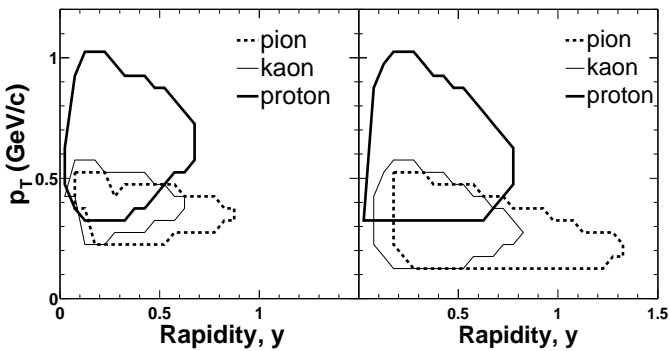}}}
\caption{Contours of the acceptance of the spectrometer as a function of transverse momentum and rapidity for pions, kaons, and protons where the raw counts per y and $p_{T}$ bin have fallen to 10\% of their maximal value. The left plot is for particles bending toward the beam pipe and the right plot is for particles bending away. The acceptance is averaged over the selected vertex range and the accepted azimuthal angle. }
\label{accept}
\end{figure}
For a given field polarity, particles of both charge signs can be reconstructed, but with different kinematic acceptances. These different acceptances are outlined in Fig.~\ref{accept}. The particles of a given charge bend in the same direction and hence have the same acceptance as do the particles of the opposite charge in the opposite polarity. 

The acceptance-corrected particle ratios are determined for each bending direction simply by the ratios of the raw counts per event for antiparticles and particles. This procedure assumes that the particle and antiparticle acceptance, tracking efficiency and kinematic distributions for each bending direction are the same over the spectrometer acceptance region for each centrality. This was verified by confirming that the opposite polarity field strength and $E_{Ring}$ fractions agree within 0.2\% and 1\% respectively. In addition, particle and antiparticle average kinematic values ($\left<p_{T}\right>$, $\left<p_{T}^{2}\right>$, $\left<y\right>$) agree within 2\% for each bending direction. Polarity dependent systematic effects that are the same for each bending direction are removed by simple averaging of the ratios measured for the two bending directions \cite{phobos200}. Examples of this include the field strength and centrality dependence.  Polarity-dependent systematic effects that are different for each bending direction, such as vertex distributions, must also be taken into account. To correct for differences in the beam orbit the data were divided into statistically independent subsets.  Differences in the vertex distribution in the z direction were accounted for by applying a z-dependent weight to the raw counts.

\begin{table*}
\begin{tabular}{|c|c|c|c|c|c|c|c|c|c|}
\toprule
Trigger & Centrality & $\left<\epsilon\right>$ & $\left<b\right>$ & $\left<N_{coll}\right>$ & $\left<\nu\right>$ & $\left<E_{Ring}^{norm}\right>$ & $\left<\pi^{-}\right>$/$\left<\pi^{+}\right>$ & $\left<K^{-}\right>$/$\left<K^{+}\right>$ & $\left<\bar{p}\right>$/$\left<p\right>$ \\ 
 Condition & \% & & & & & & & & \\ \colrule 
 dAVertex & $60-100$ & 0.20 & $7.4(1.4)$ & $2.9(1.7)$ & $2.2(1.3)$ & 0.14 & $0.995\pm0.015\pm0.017$ & $0.97\pm0.07\pm0.03$ & $0.84\pm0.04\pm0.04$  \\ 
 dAVertex & $30-60$  & 0.61 & $5.6(1.5)$ & $7.0(3.0)$ & $4.0(1.8)$ & 0.33 & $1.004\pm0.007\pm0.017$ & $0.95\pm0.03\pm0.03$ & $0.80\pm0.02\pm0.03$  \\ 
 dAVertex & $10-30$  & 0.78 & $4.0(1.5)$ & $12(3.6)$  & $6.1(1.8)$ & 0.61 & $1.008\pm0.006\pm0.017$ & $0.97\pm0.02\pm0.03$ & $0.83\pm0.02\pm0.03$  \\ 
 dAVertex & $0-10$   & 0.84 & $3.0(1.4)$ & $16(4.0)$  & $8.1(2.0)$ & 1.00 & $1.016\pm0.007\pm0.017$ & $0.97\pm0.03\pm0.03$ & $0.86\pm0.02\pm0.03$  \\ 
 dAPeriph & $60-100$ & 0.18 & $7.4(1.4)$ & $2.8(1.7)$ & $2.2(1.3)$ & 0.14 & $0.996\pm0.008\pm0.017$ & $1.02\pm0.04\pm0.04$ & $0.86\pm0.03\pm0.03$  \\ 
 dAPeriph & $30-60$  & 0.24 & $5.9(1.6)$ & $6.2(2.7)$ & $3.7(1.6)$ & 0.29 & $1.014\pm0.007\pm0.017$ & $0.97\pm0.03\pm0.04$ & $0.82\pm0.02\pm0.03$  \\ 
 \botrule
\end{tabular}	
\caption{Antiparticle to particle ratios within the acceptance for each centrality bin: $\left< \epsilon \right>$ is the trigger and vertex efficiency, $\langle b \rangle$ is the average impact parameter, $\langle N_{coll} \rangle$ is the average number of collisions, $\langle \nu \rangle$ is the number of collisions per deuteron participant,  and $\left<E_{Ring}^{norm}\right>$ is the relative multiplicity as measured by $E_{Ring}$; the numbers in parentheses represent the RMS of their respective values.  The systematic errors in these quantities are $30\%$, $20\%$, $15\%$, $10\%$ in order of increasing centrality.  The errors on the final ratios represent the statistical and point-to-point systematic errors, respectively.  The systematic scale errors are not shown. \label{table2} }
\end{table*}

Table~\ref{table4} gives a summary of the particle and event statistics used in this analysis. The raw ratios for each centrality bin are determined from the statistically weighted average of the ratios over finer subsets of the data in order to reduce systematic errors.  
\begin{table*}
\begin{tabular}{|c|c|c|c|c|c|c|c|c|c|c|c|c|c|c|c|}
\toprule
 Trigger & Centrality & \multicolumn{7}{c|}{Negative polarity} & \multicolumn{7}{c|}{Positive polarity} \\ 
 Condition & \% & Events & $\pi^{-}$ & $\pi^{+}$ & $K^{-}$ & $K^{+}$ & $\bar{p}$ & $p$ & Events & $\pi^{-}$ & $\pi^{+}$ & $K^{-}$ & $K^{+}$ & $\bar{p}$ & $p$ \\ \colrule 
 dAVertex &60-100 & 1004052 & 2787 & 13879 & 163 & 537 & 310 & 556 & 1120480 & 15320 & 3087 & 524 & 157 & 543 & 421 \\ 
 dAVertex &30-60 & 2863440 & 14963 & 74941 & 824 & 2732 & 1604 & 3494 & 3078602 & 81368 & 16164 & 2930 & 1011 & 2998 & 2282 \\ 
 dAVertex &10-30 & 2444774 & 19663 & 95755 & 1113 & 3660 & 2198 & 4558 & 2612126 & 103027 & 20641 & 3780 & 1237 & 3824 & 2903 \\ 
 dAVertex &0-10 & 1291300 & 13628 & 64357 & 803 & 2497 & 1549 & 3077 & 1378206 & 69664 & 14303 & 2675 & 864 & 2649 & 1958 \\
 dAPeriph &60-100 & 3021106 & 8255 & 40530 & 455 & 1416 & 866 & 1841 & 3222936 & 42931 & 8797 & 1509 & 461 & 1564 & 1085 \\ 
 dAPeriph &30-60 & 3053872 & 13666 & 65799 & 737 & 2353 & 1461 & 3182 & 3311696 & 71542 & 14435 & 2449 & 821 & 2707 & 2009 \\ 
\botrule
\end{tabular}
\caption{Summary of particle and event counts.\label{table4}}
\end{table*}

The systematic errors in the raw ratios were determined by examining the effects of varying the cuts used for event selection, centrality, track selection, and PID determination. There were two dominant sources of systematic errors.  The first originates from the method used to determine centrality.  Other measures of centrality lead to a point-to-point systematic error of $\pm2\%$ assigned to the kaon and proton ratios. The second major contribution originated from the dependence of the ratios on the kinematic acceptance ($p_{T}$ and $y$) over which they are measured. An error of $\pm1\%$ is assigned to the kaon and proton ratios. There are additional systematic error contributions from dead and hot spectrometer channels, spectrometer arm asymmetries, and polarity-dependent vertex corrections. For proton ratios they amounted to $0.5\%$, $1\%$, $1\%$, respectively. Electron contamination was estimated to change the $\langle \pi^{-} \rangle / \langle \pi^{+} \rangle$ ratio by less than 0.1\%.  All systematic errors on the raw ratios were added in quadrature, keeping point-to-point and scale systematic errors separate. 

The values of the ratios for detected particles will be different from those of particles produced in the collision if there is a significant yield of particles from secondary interactions and weak decays or a loss of particles due to absorption in the detector materials. Corrections resulting from additional particle yield were described in detail in refs.~\cite{phobos130,phobos200}. These corrections, and their systematic errors, are small because most of the unwanted particles can be rejected by tracking in the highly segmented silicon detectors which begin only 10~cm away from the interaction point. For the pion and kaon data, the total corrections were estimated to be less than 0.5\% and 1\%, respectively. These values are reflected in the final systematic errors of the ratios.

\begin{figure}		
\centerline{
\resizebox{0.5\textwidth}{0.35\textwidth}{
\includegraphics{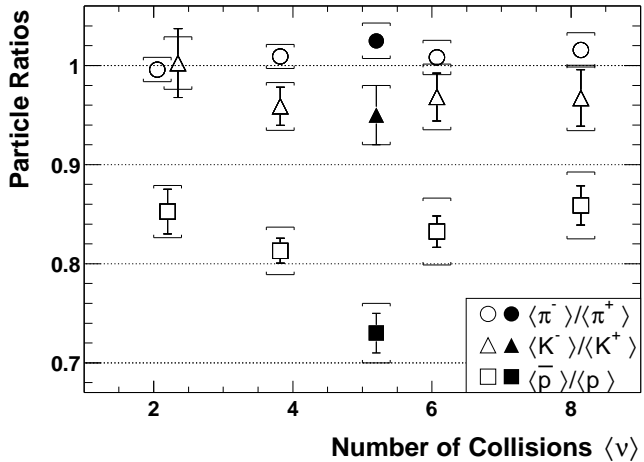}}}
\caption{Particle ratios as a function of centrality for each species. The open symbols are from d+Au collisions, the filled symbols are from central Au+Au collisions \cite{phobos200}, both at $\sqrt{s_{_{NN}}} = 200$~GeV. The brackets represent the point-to-point systematic error. The systematic scale errors are not shown.}
\label{finalratios}
\end{figure}

Corrections to the $\langle \bar{p} \rangle / \langle p \rangle$ ratio are more significant. These were determined as in \cite{phobos200} with several important differences. The dominant change to this ratio is due to the absorption correction, which arises from an asymmetry in the loss of anti-protons versus protons interacting in the beam-pipe and planes of the spectrometer. The correction value of $3.5\%\pm1.4\%$ (syst.) was determined using GEANT with two hadronic interaction packages, Gheisha and Fluka. The systematic error represents half of the difference between the two interaction packages and is a scale error for all of the proton ratios. Using HIJING events, the secondary correction for the production of protons in the beampipe and detector materials was found to be $1.6\pm0.3\%$ (stat.). The correction is larger than in the Au+Au analysis \cite{phobos200} due to the different acceptance and the reduced ability to define the event vertex. This correction is dependent on the HIJING $p/\pi$ ratio matching the true $p/\pi$ ratio. It was found that this assumption is correct at the $10\%$ level, which resulted in a systematic scale error of $\pm0.2\%$ in the correction. A feed-down correction of $-0.5\%$ accounts for the difference in the number of hyperons (primarily $\Lambda$ and $\bar{\Lambda}$) decaying to protons versus antiprotons. The reduced ability to determine an event-by-event vertex requires that a momentum dependent correction be employed. The anti-hyperon to anti-proton ratio ($\bar{\Lambda}/\bar{p}$) was estimated to be $ \approx 0.6\pm0.3$ (syst.), derived from the average of two extreme limits, the published Au+Au ratio \cite{star_lambda,phenix_lambda} and the HIJING d+Au ratio, where the model is known to underestimate strangeness production. The systematic error reflects the difference between the two values. The resulting point-to-point systematic error in the correction for the two most central bins is $\pm1\%$ and for the two peripheral bins is $\pm0.6\%$.  

Table~\ref{table2} shows the final antiparticle to particle ratios within our acceptance for the different centrality bins. The systematic scale error on the ratios has been separated from the point-to-point errors, shown in Table~\ref{table2}, allowing for a more precise determination of the centrality dependence of the ratios.  The systematic scale errors for pions, kaons and protons are $\pm0.008$, $\pm0.02$ and $\pm0.02$, respectively. For the proton ratio a large contribution to this error comes from the uncertainty in determining the absorption correction.

\begin{figure}		
\centerline{
\resizebox{0.5\textwidth}{0.35\textwidth}{
\includegraphics{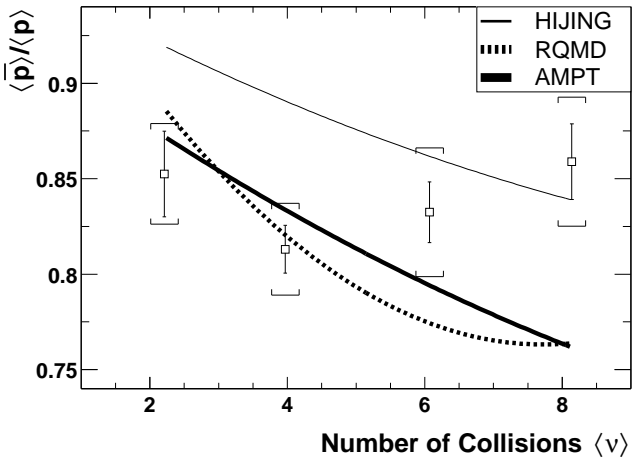}}
}
\caption{Antiproton to proton ratio comparison with models. The squares are the d+Au data at $\sqrt{s_{_{NN}}} = 200~GeV$, with brackets representing the point-to-point systematic error. The lines are fits to the model predictions of $\langle \bar{p} \rangle / \langle p \rangle$ within the acceptance.  The statistical error in the models is less than 2\%.}
\label{pbarpcompratios}
\end{figure}

Fig. \ref{finalratios} shows the final ratios for each species as a function of centrality. For clarity of presentation, the two measurements where the two data sets overlap are weighted together statistically. Within the statistical and systematic errors all the particle ratios appear to be independent of centrality in d+Au collisions. The 12\% central Au+Au ratios \cite{phobos200} are also shown in Fig. \ref{finalratios}. The value of $\left<\nu\right> \approx5.2$ for central Au+Au collisions is determined from a Glauber model calculation using HIJING. The pion and kaon ratios agree between the two systems.  This suggests that any final state interactions in Au+Au do not modify the ratio of the initially produced meson yields.  In contrast, the Au+Au proton ratio is significantly lower than the ratios in all of the centrality bins in the d+Au collisions. Additional data from Au+Au collisions \cite{star130,phobos200_QM02,phenixspectra} suggest that for less central Au+Au collisions (lower values of $\nu$) the $\langle \bar{p} \rangle/ \langle p \rangle$ ratio in our acceptance is similar to that found in the d+Au data.  

Assuming that the total proton yields are the sum of a transported component and a produced component which is equal to the antiproton yield, the value $\langle p \rangle/\langle \bar{p} \rangle~-~1$ is a measure of the relative fraction of transported protons to produced protons. A comparison of the central d+Au and Au+Au results \cite{phobos200} show that the relative fraction of transported protons in a central d+Au collision is half that observed in a central Au+Au collision, despite the larger value of $\nu$ in the central d+Au collisions. This may be evidence of collective behavior that affects baryons in Au+Au collisions and is not present in d+Au collisions.

Fig. \ref{pbarpcompratios} compares the $\langle \bar{p} \rangle / \langle p \rangle$ ratio as a function of centrality with the same ratio from HIJING \cite{hijing}, RQMD \cite{rqmd} and AMPT \cite{ampt1,ampt2}. The model outputs were passed through a simulation of the detector and the same trigger, event, and particle selection biases as used in the data analysis were applied. These models agree with the expectation of increased baryon transport with increasing $\nu$, which results in a decreasing ratio.  The ratios averaged over different centralities from the models and the data are roughly consistent.  However, suppression of the ratio with increasing centrality seen in the models is not observed in the data. 

In conclusion, the data shown in this paper provide the first information on the baryon transport in d+Au collisions at the full RHIC energy. These ratios provide constraints on current and future models dealing with baryon production and transport and thus set a baseline from which to further investigate Au+Au and other systems.

This work was partially supported by U.S. DOE grants DE-AC02-98CH10886,
DE-FG02-93ER40802, DE-FC02-94ER40818, DE-FG02-94ER40865, DE-FG02-99ER41099, 
and W-31-109-ENG-38, US NSF grants 9603486, 9722606 and 0072204, Polish KBN 
grant 2-P03B-10323, and NSC of Taiwan contract NSC 89-2112-M-008-024.

\end{document}